\documentclass[12pt]{article}
\usepackage{amssymb}
\usepackage{amsfonts}
\usepackage{amsmath}
\usepackage{amsthm}

\title{Towards experiments  to test violation of the original Bell inequality}

\author{Andrei Khrennikov\\International Center for Mathematical Modeling \\
in Physics, Engineering, Economics, and Cognitive Science\\
Linnaeus University, V\"axj\"o, Sweden \\
Irina Basieva\\ Prokhorov General Physics Institute\\  
Vavilov str. 38D, Moscow, Russia}

\begin{document}
\maketitle

\begin{abstract} The aim of this note is to attract attention of experimenters to the original Bell (OB) inequality  which was shadowed by the common consideration 
of the CHSH inequality. There are two reasons to test the OB  inequality and not the CHSH  inequality. First of all, the 
OB  inequality is  a straightforward consequence  to the EPR-argumentation. And only this inequality is related to the EPR-Bohr debate. However, 
this statement can be objected by some experts in foundations.    
 Therefore to convince experimenters 
to perform the OB-violation experiment, I prefer to concentrate on the second distinguishing feature of the 
OB  inequality which was emphasized by I. Pitowsky. He pointed out that the OB  inequality provides a higher degree 
of violations of classicality than the CHSH  inequality. For the CHSH inequality, the fraction of the quantum (Tsirelson) bound   
$Q_{{\rm CHSH}}=2\sqrt{2}$ to the classical bound $C_{{\rm CHSH}}=2,$ i.e., $F_{{\rm CHSH}}=
\frac{Q_{{\rm CHSH}}}{C_{{\rm CHSH}}}=\sqrt{2}$ is less than the fraction of the quantum bound for the OB inequality $Q_{{\rm OB}}=\frac{3}{2}$ to the 
classical bound $C_{{\rm OB}}=1,$ i.e., $F_{{\rm OB}}= \frac{Q_{{\rm OB}}}{C_{{\rm OB}}}=\frac{3}{2}.$ Thus by violating the OB inequality it is possible to approach 
higher degree of deviation from classicality. The main problem is that the OB inequality is derived under the assumption of perfect (anti-) correlations. However,  the last 
years were characterized by the amazing  development of quantum technologies. Nowadays, there exist sources producing with very high probability the pairs of 
photons  in the singlet state.  
 Moreover, the efficiency of photon detectors was improved tremendously. In any event one can start by proceeding 
with the fair sampling assumption. Another possibility is to use  the scheme of Hensen et al. experiment for 
entangled electrons. Here the detection efficiency  is very high.  
 \end{abstract}

\section{Introduction}

In his paper \cite{B_EPR} (see also \cite{B})  Bell proposed the probabilistic test based on  the EPR-argument \cite{EPR}.  
The problem under consideration can be formulated as follows.  Einstein, Podolsky, and Rosen proved that quantum mechanics (QM) is incomplete, since its formalism 
does not represent the EPR elements of reality.  Suppose one wants to construct a subquantum theory completing QM. Such a theory should match with statistical predictions 
of QM and at the same time {\it it should describe EPR's elements of reality.} Can such a theory be local? (As EPR dreamed for.)   

Bell proposed  a test based on an inequality for correlations. This inequality will be called the original Bell (OB) inequality. 
This inequality was proved under the  following crucial assumption  about coupling the Bell model with hidden variables and the
EPR elements of reality.
 
For the singlet state (as for the original EPR state), 
spin projections are EPR's elements of reality. These {\it elements of reality  are equal to measurement outcomes} (elements of reality for $S_2$ 
are measurement outcomes for $S_1).$ Hence, {\it values of variables of a subquantum theory beyond the singlet state can be identified with possible outcomes of measurements.}
Therefore, for the singlet state, subquantum and quantum correlations can be identified, see appendix for further discussion. 
 
However, this beautiful theoretical scheme supporting nonlocal hidden variable theories did not match the experimental framework of that time, since the degree
of (anti-)correlations (for the same setting on both sides) was not so high. This problem was solved by transition from the OB inequality to the CHSH inequality \cite{CHSH}
or  the similar inequalities: the CH74 inequality \cite{ CH74, CSH} or the Eberhard inequality \cite{EB} (see \cite{PSR} for comparison of these inequalities). 
Derivations of such inequalities  are not based on the assumption of  perfect (anti-) correlations. (For convenience, latter we shall compare the OB inequality only with the CHSH 
inequality, but  similar comparison can be done for other ``CHSH-like inequalities'' as the CH74 inequality and the Eberhard inequality.)  Foundational difference between the 
OB and CHSH-like inequalities is briefly discussed in appendix.

The main issue which we want to highlight in this paper is that (as was pointed by I. Pitowsky \cite{Pitowsky}) the OB  inequality provides a higher degree 
of violations of classicality than the CHSH  inequality. For the CHSH inequality, the fraction of the quantum (Tsirelson) bound   
\begin{equation}
\label{CHSHB0}
Q_{{\rm CHSH}}=2\sqrt{2}
\end{equation}
 to the classical bound $C_{{\rm CHSH}}=2,$ i.e., 
\begin{equation}
\label{CHSHB}
F_{{\rm CHSH}}=
\frac{Q_{{\rm CHSH}}}{C_{{\rm CHSH}}}=\sqrt{2}
\end{equation}

 is less than the fraction of the quantum bound for the OB inequality 
\begin{equation}
\label{OB}
Q_{{\rm OB}}=\frac{3}{2}
\end{equation}
 to the 
classical bound $C_{{\rm OB}}=1,$ i.e., 
\begin{equation}
\label{OB1}
F_{{\rm OB}}= \frac{Q_{{\rm OB}}}{C_{{\rm OB}}}=\frac{3}{2}.
\end{equation}

Thus by violating the OB inequality it is possible to approach  higher degree of deviation from classicality. 

The main problem for performing an experimental test is that the OB inequality is derived under the assumption of perfect (anti-) correlations. 
Therefore it was impossible to perform experiments to check violation of the OB inequality. 
However,  the last  years were characterized by the amazing  development of quantum technologies. 
Technological improvements led to the loophole free tests of the Bell-type inequalities\footnote{
It may be interesting for the reader that the weblinks to the video-records of the talks of the leaders of all these experimental groups accompanied with the
talks of Gregor Weihs and two talks of Philippe Grangier (at
the special session BIG EVENT: Final Bell's test, at the conference
"Quantum and Beyond", V\"axj\"o, Sweden, June 2016) can be found at the webpage of one  authors of this paper:
https://lnu.se/en/staff/andrei.khrennikov/.} (the CHSH,  Eberhard, and Clauser-Horne inequalities) \cite{B1, B2, B3}
(see also \cite{ASP}-\cite{CHR} for previous steps towards these long-aspired experiments).\footnote{As was expected by Bell, these experiments 
did not change the views of those who did not accept the conventional interpretation of experimental outputs, see, e.g., \cite{KUP}.}

 One possibility to test violation of the OB inequality  is to follow the quantum optics path initiated by Aspect \cite{ASP1}. Nowadays  there exist sources producing 
with very high probability the pairs of photons in the singlet state. 
Moreover, the efficiency of photon detectors was improved tremendously. 
Therefore  one can hope to violate the OB inequality, although this is still the real challenge,  see section \ref{CR}. In any event one can proceed under the {\it fair sampling assumption,} i.e., 
to solve first the problem of (anti-) correlations.   

Another possibility is to test the OB  inequality by using the scheme of the Hensen et al. experiment \cite{B1}. This experimental scheme does not suffer of inefficiency 
of detection. However, it seems that the quality of preparation of the singlet state is still insufficient to perform the experimental test to violate the OB  inequality, 
see section \ref{CR}.

This paper is a short review based on the results of Pitowsky \cite{Pitowsky}, Ryff \cite{ANT}, and Larsson \cite{L}. Its aim is to collect these results in one text and consider experimental 
consequences of combination of the results Ryff \cite{ANT} and Larsson \cite{L} in the light of recent tremendous achievements of modern quantum information technologies. 

In section \ref{ANTI} we present  probabilistic calculations to estimate the probability of preparation of the singlet state which is sufficient to test violation of 
the OB-inequality under the assumption of 100\%  of the detection  efficiency. Theorem 2 implies that experimenters has to be able to prepare an ensemble in which 
more than 75\% of pairs are in the singlet state, see also  Ryff's paper \cite{ANT}.  Thus the existing photon sources of high quality provide the possibility to test the OB inequality, at least under the 
assumption of fair sampling.  
 In section \ref{DE} we present probabilistic calculations to estimate  
the minimal efficiency of detection which is sufficient to test violation of 
the OB-inequality under the assumption of 100\%  fidelity in preparation of the singlet state. By Theorem 3 the efficiency of the joint detection 
should be higher than 88,9\%, see Larsson's paper \cite{L} for the original derivation of this bound. 
And finally, in section \ref{AD} we combined  the results of sections \ref{ANTI}, \ref{DE}. 
By combining 98\% level of anti-correlations with 90\% level of detection efficiency one can test violation of the OB inequality.

We remark that generalized (perturbed) Bell's inequalities  which are similar to inequalities obtained in theorems 2-4
were actively used by one of  the coauthors in foundational studies \cite{BC1}-\cite{BC4}.

Successful experimental  testing of violation of the OB inequality would be an important (although very challenging) contribution to clarification of quantum foundations.

\section{Classical and quantum bounds for the original Bell inequality}

We proceed in accordance with Bell's paper \cite{B_EPR}. 
Let $p$ be a probability measure on the space of hidden variables $\Lambda.$ (Bell used the symbol $\rho.$)
We model measurements on a pair of systems $S_1$ and $S_2$  with the aid of random variables $A_s(\lambda)$ 
and  $B_s(\lambda),$  where the parameter $s$ labels settings of measurement devices, $s=a, b,c.$ 

Consider correlations of these random variables given by the integrals: 
\begin{equation}
\label{E1}
P(a,b)= \int_\Lambda A_a(\lambda) B_b(\lambda) d p(\lambda).
\end{equation} 
It is assumed that these random variables take 
values $\pm 1$ and that the random variables corresponding to measurements on $S_1$ and $S_2$ are anti-correlated:  
\begin{equation}
\label{E1}
P(a,a) = \int_\Lambda A_a(\lambda) B_a(\lambda) d p(\lambda) =-1.
\end{equation} 
Under these assumptions Bell derived  \cite{B_EPR}, \cite{B}  the following inequality:
\begin{equation}
\label{E7}
\vert P(a,b) - P(a, c)\vert - P(b,c) \leq 1, 
\end{equation}
see also section \ref{ANTI} for details.    We call it the {\it original Bell inequality} or OB inequality.

This hidden variable model was confronted with spin measurements represented in QM by the spin operators $\sigma \cdot s.$ 
In this case $s$ is the unit vector in $\mathbf{R}^3$ representing the axis of spin projection. Thus pairwise correlations for spin operators 
 are compared with correlations for random variables. To distinguish measurements on systems $S_1$ and $S_2,$  we shall use 
 symbols $\sigma_1 \cdot a$ and $\sigma_2 \cdot b.$
 
 The OB inequality  implies that, for classical correlations,  the upper bound  $C_{{\rm OB}}$  for the expression
 $\Delta=\vert P(a,b) - P(a, c)\vert - P(b,c) $ equals to one. Now consider the the quantum case. To get perfect anti-correlations,  we proceed with the singlet state
\begin{equation}
\label{E7t}
\Psi= (\vert + -\rangle - \vert - +\rangle)/\sqrt{2}.
\end{equation}
For this state,  we have 
\begin{equation}
\label{E7a}
 P_Q(a,b)= \langle \sigma_1 \cdot a  \otimes \sigma_2 \cdot b \rangle =  - \langle a \vert b \rangle 
\end{equation}

One can find the quantum bound for the expression 
$$
\Delta_Q(a, b,c) =\vert P_Q(a,b) - P_Q(a, c)\vert - P_Q(b,c) =  \vert \langle a \vert b \rangle - \langle a \vert c \rangle \vert + \langle b \vert c \rangle.
$$ 

{\bf Theorem 1.} $Q_{{\rm OB}}= \max_{a,b,c} \Delta_Q(a, b,c)  = \frac{3}{2}.$

\medskip 

{\bf Proof.}  Under the suitable parametrization $\Delta_Q(a, b,c) $ can be represented as
\begin{equation}
\label{E7b}
\Delta_Q( \phi_1, \phi_2, \theta) = 2 \vert \sin \phi_1 \sin \phi_2 \sin \theta\vert + 1 - 2\sin^2 \phi_1. 
\end{equation}
It is easy to find that the maximal value of this function equals to 3/2. 

\medskip

Consider, for example three vectors in the same plane, $a=(1,0), b=(1/2, - \sqrt{3}/2),   c=(-1/2, - \sqrt{3}/2).$ 
Then $P(a,b)= - \langle a \vert b \rangle= - 1/2, P(a,c)= - \langle a \vert c \rangle=  1/2, P(b,c)= - \langle b \vert c \rangle= - 1/2.$ 
Hence, $\Delta_Q(a,b,c)= 3/2.$

Hence, we proved the equality (\ref{OB1}), $F_{{\rm OB}}=3/2.$ 

I. Pitowsky \cite{Pitowsky} presented the same argument by using a slight modification  of the OB inequality (\ref{E7}). 

\section{Original Bell inequality: taking into account imperfection of anti-correlations}
\label{ANTI}

Here we proceed in Bell's framework based on classical probability  under the assumption 
 that the random variables corresponding to measurements on $S_1$ and $S_2$ are anti-correlated.  
As Bell pointed out, this is possible only if the following equality holds 
\begin{equation}
\label{E2}
A_a(\lambda) = - B_a(\lambda) ,
\end{equation} 
except a set of measure zero. Bell derived inequality (\ref{E7}) under this assumption of perfect (up to measure zero) anti-correlation. It is easy to modify 
this equality under assumption of imperfect anti-correlations. Here we follow the original paper \cite{ANT}, but we proceed 
in measure theoretic framework. Using of the frequentist approach (as in paper \cite{ANT}) have been objected by a few authors, see, e.g., \cite{BC1}.

 Suppose that, for each $a,$ there exists a subset  $\Lambda_a$ of $\Lambda$ such that eq. (\ref{E2}) holds 
for all $\lambda$ from $\Lambda_a$ and, for the set $\Lambda_a^\prime= \Lambda \setminus \Lambda_a,$ we have: 
\begin{equation}
\label{E3}
p(\Lambda_a^\prime) \leq  \epsilon.
\end{equation} 
Since random variables are dichotomous, on the set $\Lambda_a^\prime$    
\begin{equation}
\label{E2X}
A_a(\lambda) = B_a(\lambda) .
\end{equation} 

Now on $\Lambda_b$ we have: 
$$
A_a(\lambda) B_c(\lambda) - A_a(\lambda) B_b(\lambda) = 
 A_a(\lambda) A_b(\lambda)A_b(\lambda)  B_c(\lambda) + A_a(\lambda) A_b(\lambda)
 $$
 $$
 = A_a(\lambda) A_b(\lambda) [1 + A_b(\lambda) B_c(\lambda)] .
 $$
On $\Lambda_b^\prime$ we have: 
$$
A_a(\lambda) B_c(\lambda) - A_a(\lambda) B_b(\lambda) = 
$$
$$
 A_a(\lambda) A_b(\lambda)A_b(\lambda)  B_c(\lambda) + A_a(\lambda) A_b(\lambda) - 2 A_a(\lambda) A_b(\lambda)
 $$
 $$
 = A_a(\lambda) A_b(\lambda) [1 + A_b(\lambda) B_c(\lambda)]  - 2 A_a(\lambda) A_b(\lambda).
 $$
Thus 
$$
P(a,c) - P(a,b)= \int_{\Lambda} A_a(\lambda) A_b(\lambda) [1 + A_b(\lambda) B_c(\lambda)] d p(\lambda) - 
2 \int_{\Lambda_b^\prime}  A_a(\lambda) B_b(\lambda) d p(\lambda).
$$
Hence we proved the following theorem:   

\medskip

{\bf Theorem 2.} [Ryff \cite{ANTI}] (Generalization of the OB inequality for imperfect anti-correlations).   
{\it Under assumption  (\ref{E3}) the following inequality for classical 
correlations holds: }
\begin{equation}
\label{E6}
\vert P(a,b) - P(a, c)\vert - P(b,c)  \leq 1+    2 \epsilon.  
\end{equation}  
\medskip

By introducing the parameter $\gamma=1-\epsilon$ we write (\ref{E6}) as  
\begin{equation}
\label{E6}
\vert P(a,b) - P(a, c)\vert - P(b,c)  \leq 3  -  2 \gamma.  
\end{equation}  

By theorem 1 we have the inequality:  $3 - 2 \gamma < 1, 5,$ i.e., $ \gamma > 3/4 = 0, 75.$ Thus to be able to test properly the OB inequality one has to be able to produce
an ensemble of pairs of quantum systems in which the percentage of precisely (anti-) correlated pairs will be higher than 
\begin{equation}
\label{E8}
\gamma  =75 \%.
\end{equation}  

\section{Original Bell inequality: taking into account the detection efficiency}
\label{DE}

For the OB inequality, the issue of the detection efficiency was studied in details by J.-A. Larsson \cite{L}. Here we present similar consideration, but in slightly different form
which is consistent with the above presentation of the role of imperfection of correlations. Denote the set of hidden variables for which the pair $A_a(\lambda), B_b(\lambda)$ 
is detected by the symbol $\Gamma_{ab}.$ The main parameter of  the experimental interest is the  
probability of joint detection of a pair,  $p(\Gamma_{ab}).$ 
For simplicity of considerations, we assume that this probability does not depend on the pair of settings, i.e., 
\begin{equation}
\label{E9tt}
\eta \equiv p(\Gamma_{ab}).
\end{equation}
Then correlation conditioned on the pairwise 
detection is given by 
\begin{equation}
\label{E9}
\tilde{P}(a,b)= \frac{1}{\eta} \int_{\Gamma_{ab}}  A_a(\lambda) B_b(\lambda) d p(\lambda).
\end{equation}  

\medskip

{\bf Theorem 3.} {\it Under the assumptions of 100\% perfect anti-correlations and set-independent joint detection efficiency, 
see (\ref{E9tt}) , the following  OB inequality for detectable correlations holds:}
\begin{equation}
\label{E10}
\vert \tilde{P}(a,b) -   \tilde{P}(b,c)\vert - \tilde{P}(a,c) \leq  \frac{(4-3 \eta)}{\eta}.  
\end{equation}

\medskip

{\bf Proof.} From (\ref{E9}) we get: 
$
 \eta \tilde{P}(a,b)= P(a, b) - u_{ab},$ where  
$$
u_{ab}= \int_{\Lambda \setminus \Gamma_{ab}}  A_a(\lambda) B_b(\lambda) d p(\lambda)
$$ 
and, hence,
 $\vert u_{ab} \vert \leq (1-\eta). $
 We have
$$
  \eta  \Big( \vert \tilde{P}(a,b) -   \tilde{P}(b,c)\vert  - \tilde{P}(a,c)\Big)
$$
$$
\leq  \vert P(a, b) - P(b,c) \vert - P(b,c) + 3(1-\eta)  \leq  1 + 3(1-\eta). 
$$
By dividing both sides of this inequality by $\eta$ we obtain (\ref{E10}). 

\medskip

To be able to violate inequality (\ref{E10}), experimenter has to have sufficiently high the detection efficiency, such that 
$\frac{(4-3 \eta)}{\eta} < \frac{3}{2},$ i.e., $\eta > 8/9= 0.889.$ Thus  the {\it  efficiency of the joint detection 
should be higher than 88,9\%.} This result coincides with the corresponding result from \cite{L}, 
p. 57. Thus the detection efficiency should be higher than in the experimental tests for the CHSH inequality \cite{GM}, \cite{L}. 

\section{Original Bell inequality: taking into account imperfection of anti-correlations and  the detection efficiency}
\label{AD}

{\bf Theorem 4.} {\it Under the assumptions (\ref{E9tt}) and (\ref{E3}), the following experimentally testable 
version of the OB inequality holds: }
\begin{equation}
\label{E10T}
\vert \tilde{P}(a,b) -   \tilde{P}(b,c)\vert - \tilde{P}(a,c) \leq  \frac{(4 + 2 \epsilon -3 \eta)}{\eta}.  
\end{equation}

\medskip

{\bf Proof.} We have
$$
  \eta \Big(\vert \tilde{P}(a,b) -   \tilde{P}(b,c)\vert  - \tilde{P}(a,c)\Big)   \leq  \vert P(a, b) - P(b,c) \vert - P(b,c) + 3(1-\eta) 
$$
$$
 \leq  1 + 2\epsilon + 3(1-\eta). 
$$
\medskip

To be able to violate inequality (\ref{E10T}), experimenter has to have  sufficiently high anti-correlations and  the detection efficiency, such  that
$\frac{4 + 2\epsilon - 3\eta}{\eta} < \frac{3}{2}.$ It is convenient to introduce a new parameter $\kappa= 1-\epsilon.$ Then the generalized OB inequality has the form:
\begin{equation}
\label{E10y}
\vert \tilde{P}(a,b) -   \tilde{P}(b,c)\vert - \tilde{P}(a,c) \leq  \frac{(6 - 2 \gamma -3 \eta)}{\eta}.  
\end{equation}
and the condition for possible violation can written as
\begin{equation}
\label{E10pa}
4\gamma + 9 \eta > 12.
\end{equation}

For example, let $\gamma= 0.98.$ Then $\eta>  0.9.$ Thus by approaching  the 98\% -level  of anti-correlations and the 90\% -level the detection efficiency experimenter can  test the OB inequality.  

\section{Concluding remarks}
\label{CR}

The modern quantum technology provides the sources producing  photons in the singlet state with very high probability, up to 98\% of generated ensemble of pairs. From this viewpoint 
it is promising  to perform the experimental test for the  OB inequality by using  entangled photons, cf. with experiments \cite{B2, B3} to violate the CHSH-like  inequalities 
(the Eberhard and CH inequalities). However, as we have
seen tests for  the OB inequality demand  higher detection efficiency than tests for the CHSH inequality.We remind (see also \cite{PSR} for discussion) that the detection efficiency is not reduced 
to the efficiency of photo-detectors. Although nowadays there are available photo-detectors having the efficiency close to 100\%, this does not solve the problem of the detection efficiency. A weak 
element of the experimental setup  based on quantum optics  is polarization beam splitter, where one can lose 8-13\% of photons. This lost can play the crucial role in attempts to lift the 
detection efficiency from  83\%     \cite{GM, L} in the tests for the CHSH inequality to approximately 90\% in the planned experimental test for the OB inequality. 

It may be reasonable to proceed under  the assumption of {\it fair sampling}. And such a project seems to be realizable.         

If one want to proceed without the fair sampling assumption, then it is very promising to test violation if the OB inequality by  
 using entangled electron spins, i.e., the scheme of Hensen et. al. \cite{B1} 
experiment which was done for the CHSH inequality. As was reported in \cite{B1}, the parameter $\gamma$ in inequality (\ref{E10y})
can be selected as $\gamma\approx 0.92.$  It exceeds the bound $\gamma=0, 75,$ see eq. (\ref{E8}) . 
So, it seems that such an experiment can be performed already today.

 {\bf Acknowledgments.} 
The authors would like to thank A. Cabelo, E. Dzhafarov, G. Jager, J.-A. Larsson, S. Polyakov, K. Svozil, G. Weihs, S. Glancy; F. De Martini  for critical discussions and knowledge transfer; 
in particular, we are thankful to G. Jaeger for pointing to Pitowsky's paper 
\cite{Pitowsky}. The second coauthor (AKH) was supported by the EU-project "Quantum Information Access and Retrieval Theory" (QUARTZ), Grant No. 721321.  
   
\section*{Appendix}

The original Bell project can be formulated as following: 
\begin{itemize}

\item Einstein, Podolsky, and Rosen proved the existence of elements of reality  (for the very special state).
\item This implies that QM is not complete and it has to be considered as emergent from some theory with hidden variables.
\item Einstein, Podolsky, and Rosen expected that such a theory would be local. (They did not construct such a theory, but they dreamed for it.)
\item  Bell's message based on violation of the OB inequality by (theoretical) quantum correlations:   unfortunately,  
EPR realism is not compatible with locality.
\end{itemize}
 We cite Bell \cite{B_EPR}, p. 195:   
``Since we can predict in advance the result of measuring any chosen component
of $\sigma_2 ,$ by previously measuring the same component of $\sigma_1,$  it follows that the result of any such
measurement must actually be predetermined. Since the initial quantum mechanical wave function does not
determine the result of an individual measurement, this predetermination implies the possibility of a more
complete specification of the state.'' Thus Bell's study was aimed to check reazability the EPR project: to construct a subquantum 
model which would match with statistical predictions of QM and at the same time describe the  EPR elements of reality. 

We remark that one could respond straightforwardly to EPR's argument by saying that measurement of the system without disturbance 
is impossible because a faster-than-light signal can move from $S_1$  to $S_2.$ 
This response EPR would treat as non-physical. And this is the important point.\footnote{We can mention Bohr's response to the EPR paper \cite{BR0}. However, it seems that 
Bohr did not understand the EPR argument. In any event his reply does not explain the origin of perfect correlations.}  

However, Bell proved (see  \cite{B_EPR}, p. 199): : 
``In a theory in which parameters are added to quantum mechanics to determine the results of individual
measurements, without changing the statistical predictions, there must be a mechanism whereby the setting
of one measuring device can influence the reading of another instrument, however remote. Moreover,
the signal involved must propagate instantaneously, so that such a theory could not be Lorentz invariant.''
Thus he treated violation of the OB inequality as the proof of nonlocality of any theory with hidden variables.

Now, we point to the crucial connection between the EPR argument and the OB inequality.  For the singlet state (as for the original EPR state), 
spin projections are EPR's elements of reality. These elements per definition are equal to measurement outcomes (elements of reality for $S_2$ 
are measurement outcomes for $S_1).$ Hence, values of variables of a subquantum theory beyond the singlet state can be identified with possible outcomes of measurements.
Therefore, for the singlet state, subquantum and quantum correlations can be identified.
 
There are no reasons to assume this for non-singlet state. Therefore CHSH-like projects which do not straightforwardly  based on the perfect (anti-)correlations can be objected 
from the viewpoint that there is no reason to identify the values of subquantum and quantum variables and hence subquantum and quantum correlations. Subquantum correlations 
satisfy CHSH-inequality, but quantum correlations violate it. (In particular, this was the viewpoint of L. De Broglie, see \cite{AFTER} for details and references.). 
By rephrasing Bell  we can say ``that what is proved, by impossibility proofs, is lack of imagination'' of possible couplings beween subquantum and quantum correlations (cf. \cite{BY}).

Therefore it is important to perform experimental test for the OB inequality. This and only this test would imply that the issue of nonlocality has to be considered seriously.

\end{document}